\documentclass[11pt,twoside]{article}

\usepackage{asp2014}

\aspSuppressVolSlug
\resetcounters

\begin{document}

\title{TOPCAT Corner Plot}

\author{Mark~Taylor}
\affil{H.~H.~Wills Physics Laboratory, Tyndall Avenue,
       University of Bristol, UK;
       \email{m.b.taylor@bristol.ac.uk}}

\paperauthor{Mark~Taylor}{m.b.taylor@bristol.ac.uk}{0000-0002-4209-1479}{University of Bristol}{School of Physics}{Bristol}{Bristol}{BS8 1TL}{U.K.}



\begin{abstract}
TOPCAT is a desktop GUI tool for working with tabular data
such as source catalogues.  Among other capabilities it provides
a rich set of visualisation options suitable for interactive
exploration of large datasets.
The latest release introduces a Corner Plot window which displays
a grid of linked scatter-plot-like and histogram-like plots for
all pair and single combinations from a supplied list of coordinates.
\end{abstract}



\section{Introduction}

TOPCAT\footnote{\url{http://www.starlink.ac.uk/topcat/}}
\citep{2005ASPC..347...29T} is an established
desktop GUI application
for working with tabular data.
It has many capabilities including
manipulation of data and metadata,
Virtual Observatory access, and
crossmatching,
but one of its particular strengths is to enable
interactive exploratory analysis of large tables.
Such tables may have millions of rows and hundreds of columns.

Exploration of such bulky and high-dimensional data is facilitated
in TOPCAT by provision of a number of highly configurable and
responsive interactive
visualisation windows, focussed on presentation of
point clouds over a wide range of densities.
Features include scatter plots in 1, 2 and 3 dimensions,
the ability to label points with shapes, vectors, error bars
and text, extensive options for colouring points by
density or additional coordinates,
and the ability to plot arbitrary combinations of table columns
using a powerful expression language.

The most recent release, v4.9, introduces a new ``Corner''
plot window,
which presents a grid of scatter-plot-like and histogram-like
plots for all pair and single combinations of a supplied list of
coordinates.
This provides a simultaneous view of all the 2-d and 1-d projections
of a high-dimensional dataset,
which can be a valuable aid for identifying interesting features
in complex data.
An example is shown in Figure~\ref{P113_screenshot}.
This plot type is also available as the command
{\tt plot2corner} in STILTS \citep{2006ASPC..351..666T},
the command-line counterpart of TOPCAT.

This type of visualisation is not novel;
it has been used since the 1980s \citep{cleveland93} under the names
{\em ``Scatter Plot Matrix''},
{\em ``SPLOM''},
{\em ``Pairs Plot''} and
{\em ``Corner Plot''},
and graphics packages producing them
exist for instance in Python, R and IDL.
Its adoption here
benefits from the performance, interactive features
and ease of use provided by the existing application GUI,
and gives TOPCAT users a new tool for exploring high-dimensional data.

\articlefigure[height=0.9\textheight]
              {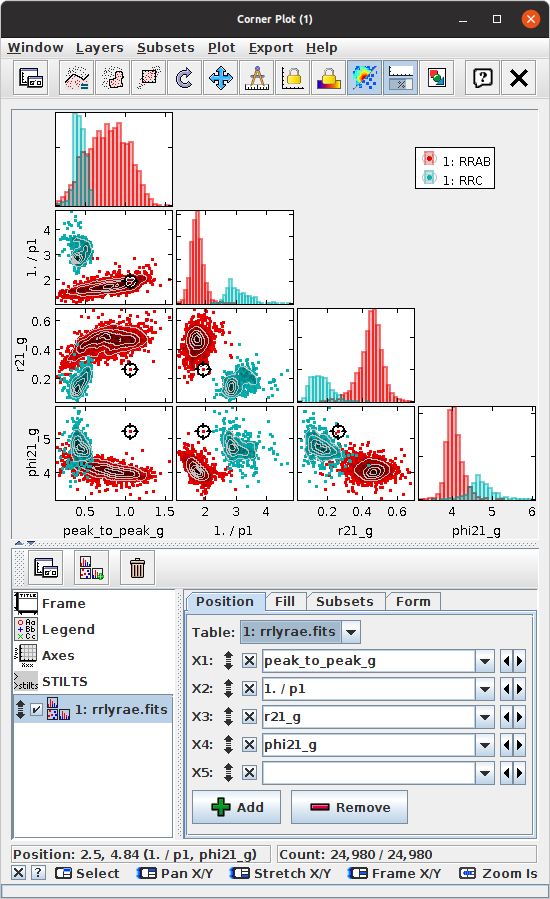}
              {P113_screenshot}
              {Corner Plot Window showing variability relationships
               between two different classes of RR-Lyrae stars;
               data is from Gaia DR1 \citep{2016A&A...595A.133C}.}

\section{Features}

The Corner Plot shares its features with the other
TOPCAT visualisation windows:
\begin{description}
\item[Ease of use:]
      coordinates can be added, deleted and reordered
      with a few mouse clicks
\item[Interactivity:]
      each panel can be panned and zoomed in one or two dimensions
      easily using the mouse,
      and the other panels in the same row/column adjust
      to keep the axes aligned
\item[Linked Views:]
      graphical selections and point highlighting on one panel
      are immediately visible in the other panels
      (and in other windows)
\item[Overplotting:]
      multiple datasets, selections and
      plot types can appear in each panel
\item[Calculations:]
      coordinates can be existing columns or
      algebraic expressions based on them
\item[Scalability:]
      multipanel displays for millions of rows
      are quite responsive even on modest hardware
\item[Flexibility:]
      many plotting and shading options are available
      alongside simple scatter plots and histograms:
      transparency, weighted density maps, contours, KDEs, text labels,
      etc
\item[Configurability:]
      interactive controls can specify  lower/upper/full panel matrix,
      logarithmic or inverted axes, axis labels, grid line drawing,
      colour map adjustments, etc
\end{description}

The linked views, especially in combination with interactive graphical
region selection (blob drawing),
make it very easy to identify regions,
populations or points that are apparent in one
coordinate plane and see where they appear in the other displayed planes.

To facilitate setting up plots with many panels,
an extra ``Fill'' control is provided
which allows easy selection of coordinates,
including an option to use all pair differences (or ratios)
from a selection of columns,
for instance all pairwise colours from a list of
magnitudes (or fluxes)

\section{Implementation}

Assembling multiple plot panels in this way
is not conceptually difficult,
and the TOPCAT codebase already provided
most of the data handling infrastructure required,
as well as some support for multi-panel plots.
However, implementing a plot type with multiple panels
derived from a single set of input coordinates broke
some of the assumptions in the existing plotting framework.
This meant that significant code restructuring was required,
especially for the graphical and command-line user interfaces,
so implementation was quite time-consuming.

\section{Future Work}

The new corner plot is working quite well, but not perfectly.
In particular the axis labelling developed for other plot types
can become overcrowded and imperfectly aligned.
Some performance improvements could be made by concurrent plotting
of multiple panels and eliminating some duplicated work.
Usage in the field may come up with other bugs, complaints or suggestions.
If this plot type proves popular,
improvements may be made in future, in line with user feedback.
Other possibilities for multi-panel plots may also be investigated.

\acknowledgements This work was funded by the UK's
Science and Technology Facilities Council (STFC).

\bibliography{P113}


\end{document}